\documentclass[9pt, conference]{IEEEtran}
\IEEEoverridecommandlockouts

\usepackage{cite}
\usepackage{amsmath,amssymb,amsfonts}
\usepackage{graphicx}
\usepackage{textcomp}
\usepackage{xcolor}
\usepackage{xurl}
\usepackage{hyperref}

\usepackage{comment, url, algorithm, algpseudocode, makecell}

\def\BibTeX{{\rm B\kern-.05em{\sc i\kern-.025em b}\kern-.08em
    T\kern-.1667em\lower.7ex\hbox{E}\kern-.125emX}}
\begin{document}

\title{FlashSR: One-step Versatile Audio Super-resolution via Diffusion Distillation\\
\thanks{This work was supported by the National Research Foundation of Korea (NRF) grant funded by the Korea government (MSIT) (No. RS-2023-00222383).}
}

\author{\IEEEauthorblockN{Jaekwon Im}
\IEEEauthorblockA{\textit{Graduate School of Culture Technology} \\
\textit{KAIST}\\
Daejeon, South Korea \\
jakeoneijk@kaist.ac.kr}
\and
\IEEEauthorblockN{Juhan Nam}
\IEEEauthorblockA{\textit{Graduate School of Culture Technology} \\
\textit{KAIST}\\
Daejeon, South Korea \\
juhan.nam@kaist.ac.kr}
}

\maketitle

\begin{abstract}
Versatile audio super-resolution (SR) is the challenging task of restoring high-frequency components from low-resolution audio with sampling rates between 4kHz and 32kHz in various domains such as music, speech, and sound effects. Previous diffusion-based SR methods suffer from slow inference due to the need for a large number of sampling steps. In this paper, we introduce FlashSR, a single-step diffusion model for versatile audio super-resolution aimed at producing 48kHz audio. FlashSR achieves fast inference by utilizing diffusion distillation with three objectives: distillation loss, adversarial loss, and distribution-matching distillation loss. We further enhance performance by proposing the SR Vocoder, which is specifically designed for SR models operating on mel-spectrograms. FlashSR demonstrates competitive performance with the current state-of-the-art model in both objective and subjective evaluations while being approximately 22 times faster.
\end{abstract}

\begin{IEEEkeywords}
audio super-resolution, bandwidth extension, diffusion probabilistic model, generative model.
\end{IEEEkeywords}

\section{Introduction}
\label{sec:intro}
Audio super-resolution (SR) is the process of reconstructing high-frequency content from low-resolution audio to produce high-resolution audio, which is expected to enhance user listening experiences. This technique can be applied to improve the quality of various audio types, including narrowband telephone audio, historical recordings, and output of audio generative models \cite{audiosr}. Deep learning-based SR methods are widely used due to their superior performance over traditional methods. While early approaches primarily focused on reconstructing 16kHz waveforms \cite{li2015dnn, hou2020}, recent works have increasingly aimed at generating 48kHz waveforms \cite{nuwave, bwallyou}. Most previous studies evaluated their work in confined settings, typically using a fixed upsampling ratio and focusing on limited domains such as speech or music \cite{musicsr}.

AudioSR \cite{audiosr} is the first SR model operating across the general audio domain, encompassing music, speech, and sound effects. It is also capable of upsampling audio with arbitrary sampling rates between 4kHz and 32kHz to 48kHz, making it suitable for real-world applications. AudioSR is composed of a latent diffusion model (LDM) \cite{stablediffusion}, a variational autoencoder (VAE) \cite{vae}, and HiFi-GAN \cite{hifigan}. The process begins with the VAE encoding the low-resolution mel-spectrogram into a latent representation. The LDM then predicts the corresponding latent for the high-resolution mel-spectrogram. The predicted latent is transformed into a mel-spectrogram using the VAE decoder. Finally, HiFi-GAN converts the mel-spectrogram into high-resolution waveforms. Despite its strong performance, AudioSR has several limitations. First, it requires lower-frequencies replacement (LFR) post-processing, where the lower frequency part of the generated mel-spectrogram and the STFT spectrogram are replaced with the original inputs. This process ensures consistency in low-frequency information and compensates for the limited perceptual quality of audio generated by HiFi-GAN. However, we found that this post-processing often makes unnatural connections between the generated high-frequency components and the input low-frequency components, leading to unpleasant artifacts. Additionally, despite calculating the cutoff frequency energy ratio to match the energy levels between the two components, the final output audio sometimes shows discrepancies in energy between the high-frequency and low-frequency components. Second, since the diffusion model relies on iterative denoising, it suffers from slow inference speed. The latent diffusion model of AudioSR generates samples by the DDIM sampler \cite{ddim} in 50 steps with classifier-free guidance (CFG) \cite{cfg}, requiring 100 Neural Function Evaluations (NFEs). Our test showed that AudioSR takes approximately 8.19 seconds to generate 5.12 seconds of audio on a single A6000 GPU, which is well below real-time performance. This limitation can restrict its practical applications. 

Reducing the number of sampling steps of diffusion models is an active research area. Improved numerical solvers \cite{ddim, dpmsolverplusplus} have been developed to perform sampling in fewer steps, but in most cases, these solvers still require more than 10 NFEs. Another approach is the distillation method \cite{progressivedistill, flashdiffusion}, where a pre-trained teacher network is used to train a student network to generate samples in fewer steps. Progressive distillation \cite{progressivedistill} uses a repetitive process, enabling models to generate images in as few as 4 steps while maintaining high quality. Flash Diffusion \cite{flashdiffusion} proposed a robust distillation method for a few steps image generation and evaluated it on various image generation tasks. Although diffusion distillation has proven successful in the image domain, its application in the audio domain \cite{consistencyTTA, musicconsistency, liu2024audiolcm, soundctm, prodiff, comospeech} remains relatively unexplored.

\begin{figure*}[htbp]
\centerline{\includegraphics[width=\textwidth]{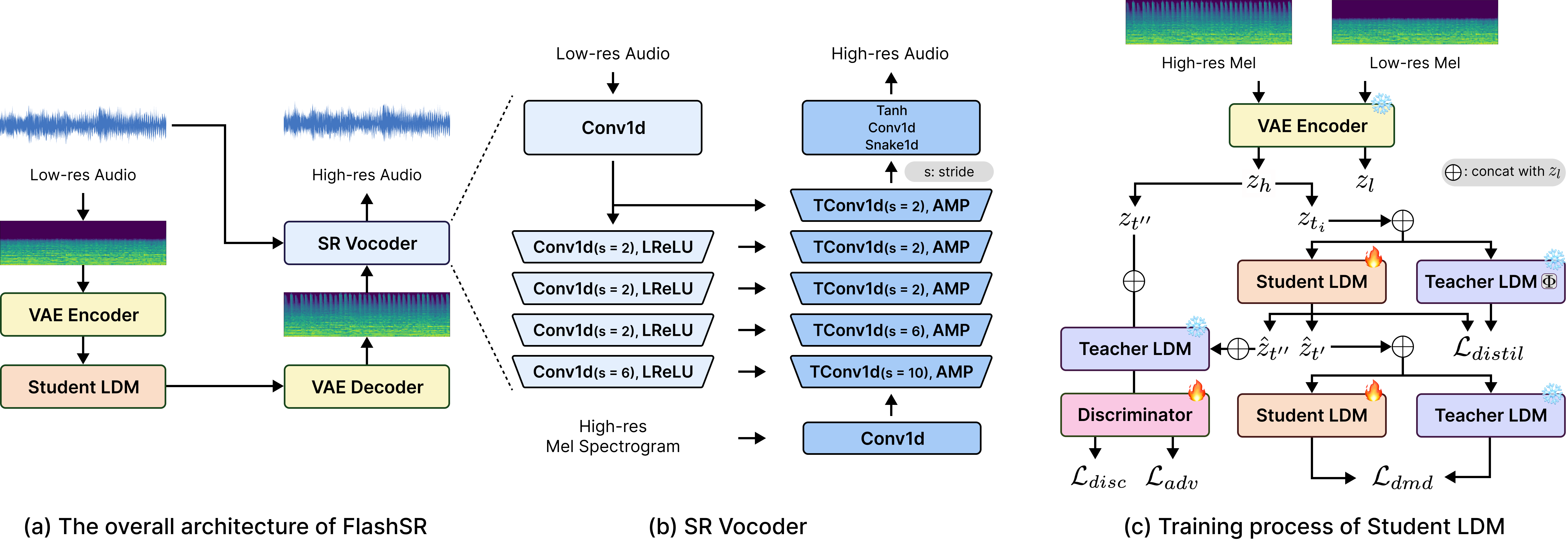}}
\caption{Overview of FlashSR.}
\label{fig:overallarchitecture}
\end{figure*}

In this paper, we introduce FlashSR, a one-step diffusion model for audio super-resolution that can upsample various types of audio, including music, speech, and sound effects, from any sampling rate between 4kHz and 32kHz to 48kHz. FlashSR only takes approximately 0.36 seconds to produce 5.12 seconds of audio on a single A6000 GPU, making it about 22 times faster than AudioSR \cite{audiosr}. This significant improvement is achieved through two major modifications to the AudioSR framework. First, we adapt the distillation method from Flash Diffusion \cite{flashdiffusion}, enabling the LDM to generate samples with only 1 NFE. To the best of our knowledge, FlashSR is the first system to apply diffusion distillation to the audio super-resolution task. Second, we propose SR Vocoder, which further improves inference speed by eliminating the need for LFR post-processing. Given a mel-spectrogram and a low-resolution waveform with an arbitrary sample rate between 4kHz and 32kHz, SR Vocoder effectively generates a high-resolution waveform having a natural connection between high-frequency and low-frequency components. We show that FlashSR performs comparably to AudioSR in both objective and subjective evaluations despite its significantly faster inference speed. Audio examples of FlashSR are available online\footnote{\url{https://jakeoneijk.github.io/flashsr-demo}}.

\section{Method}
\label{sec:methodology}

Figure \ref{fig:overallarchitecture} (a) illustrates the overall architecture of FlashSR. Let $x_h \in \mathbb{R}^{s \cdot h}$ and $x_l \in \mathbb{R}^{s \cdot l}$ represent high-resolution and low-resolution audio, respectively, where $s$ denotes the duration in seconds, and $h$ and $l$ are the sample rates of the high and low-resolution audio, with $h > l$. Their mel-spectrograms are represented by $X_h \in \mathbb{R}^{F \times T}$ and $X_l \in \mathbb{R}^{F \times T}$, where $F$ corresponds to the number of mel frequency bins and $T$ corresponds to the number of time frames. Note that $X_h$ and $X_l$ have the same dimensions because $x_l$ is resampled using a windowed sinc interpolation to match the size of $x_h$ before the mel-spectrogram is calculated. For LDM working on a VAE latent space, each mel-spectrogram is compressed into $z_h \in \mathbb{R}^{C \times \frac{T}{r} \times \frac{F}{r}}$ and $z_l \in \mathbb{R}^{C \times \frac{T}{r} \times \frac{F}{r}}$ by the pre-trained VAE encoder, where $C$ denotes the channel size and $r$ denotes the compression level. The Student LDM estimates $z_h$ from $z_l$, and the predicted latent  $\hat{z}_h$ is converted into a high-resolution mel-spectrogram $\hat{X}_h$ by a pre-trained VAE decoder. Using $\hat{X}_h$ and $x_l$, SR Vocoder then generates the high-resolution waveform $\hat{x}_h$. In Section \ref{subsec:flash}, we describe the training process for the  Student LDM. In Section \ref{subsec:srvocoder}, we introduce the SR Vocoder in detail.

\subsection{Flash Diffusion}
\label{subsec:flash}
    
\begin{algorithm}[!t]
\caption{Training  Student LDM}\label{alg:flashdiffusion}
\begin{algorithmic}
\Require dataset $\mathcal{D}$, 
\Require trained  Teacher LDM $f_{\eta}$,  Student LDM $f_{\theta}$ \Require discriminator $d_{\nu}$
\Require ODE solver $\Phi$, guidance scale $\omega$
\Require timesteps distribution $\pi(t)$
\Require adv loss weight $\lambda_{adv}$, dmd loss weight $\lambda_{dmd}$

\State $\theta \gets \eta$ \Comment{Init student weights with teacher weights}
\Repeat
\State ($z_h$, $z_l$) $\sim \mathcal{D}$
\State $t_{i} \sim \pi(t)$, $\epsilon \sim \mathcal{N}(0,I)$
\State $z_{t_i} \gets \alpha_{t_i}z_{h} + \sigma_{t_i}\epsilon$
\State $\hat{z}_{t_0} \gets \alpha_{t_i}z_{t_i}- \sigma_{t_i} f_{\theta}(z_{t_{i}}, t_{i},z_l)$
\For{$j = i - 1$ to $0$}
\State $\hat{v}_{t_{j+1}} = \omega f_{\eta}(z_{t_{j+1}}, t_{j+1},z_l) + (1-\omega) f_{\eta}(z_{t_{j+1}}, t_{j+1})$
\State $z_{t_j} \gets \Phi(\hat{v}_{t_{j+1}},  t_{j+1}, z_{t_{j+1}})$
\EndFor
\State $\mathcal{L}_{\theta} \gets \mathcal{L}_{distil}(\hat{z}_{t_0}, z_{t_0}) + \lambda_{dmd}\mathcal{L}_{dmd}(\hat{z}_{t_0},z_l, f_{\eta}, f_{\theta}) $
\State \indent $+ \lambda_{adv}\mathcal{L}_{adv}(\hat{z}_{t_0}, z_l, d_{\nu}, f_{\eta})$
\State Update $\theta$ with gradient $\nabla_{\theta}\mathcal{L}_{\theta}$
\State $\mathcal{L}_{\nu} \gets \mathcal{L}_{disc}(\hat{z}_{t_0}, z_h, z_l, d_{\nu}, f_{\eta})$
\State Update $\nu$ with gradient $\nabla_{\nu}\mathcal{L}_{\nu}$

\Until{convergence}
\end{algorithmic}
\end{algorithm}

We use AudioSR \cite{audiosr} as the  Teacher LDM $f_{\eta}$ and train the  Student LDM $f_{\theta}$, which has the same architecture as the teacher, utilizing the diffusion distillation method proposed in Flash Diffusion \cite{flashdiffusion}. $\eta$ and $\theta$ are the parameters of the  Teacher LDM and the  Student LDM, respectively, with only $\theta$ being updated during training. Since Flash Diffusion is compatible with LoRA \cite{lora}, we apply LoRA to the student model to decrease training parameters and enhance training efficiency. The training process for the  Student LDM is outlined in Algorithm \ref{alg:flashdiffusion}. The diffused sample $z_{t_i}$ is obtained by applying the forward diffusion process \cite{ddpm}, specified by $\alpha_{t_i}$ and $\sigma_{t_i}$ from the diffusion noise schedule, to $z_h$. The diffusion timestep $t_i$ is sampled from a mixture of Gaussian distributions $\pi(t)$, designed to prioritize the timesteps relevant to few-step data generation. Given $z_{t_i}$, the  Student LDM estimates $\hat{z}_{t_0}$ in $z_{h}$-space. Following AudioSR, the  Student LDM is parameterized to predict $v$ \cite{progressivedistill}. As illustrated in Figure \ref{fig:overallarchitecture} (c), the Student LDM is trained with three objectives, which are explained in the following.

\noindent \textbf{Distillation loss:} Instead of using $z_h$, the distillation loss target is derived from the  Teacher LDM. The denoised latent $z_{t_0}$ is generated by the  Teacher LDM using an ODE solver with classifier-free guidance. We found that sampling a guidance scale proposed in Flash Diffusion leads to more unstable training for our task. Therefore, we use a fixed guidance scale $\omega$ to generate $z_{t_0}$. By using $z_{t_0}$ as the target, the  Student LDM can learn the data distribution modeled by the teacher:
\begin{equation}
    \mathcal{L}_{distil} = \mathbb{E}_{(z_h,z_l),t,\epsilon} \left[ \lVert \hat{z}_{t_0} - z_{t_0} \rVert^{2} \right]
\end{equation}

\noindent \textbf{Distribution Matching Distillation (DMD) loss:} The DMD loss \cite{dmdloss} enables the Student LDM to learn the  Teacher LDM's knowledge at the distribution level. This is achieved by minimizing the Kullback–Leibler (KL) divergence between $p_{\theta}$ and $p_{\eta}$, the distributions of samples from the Student and  Teacher LDM, respectively.
\begin{equation}
\begin{split}
    \mathcal{L}_{DMD} &= D_{KL}(p_{\theta}||p_{\eta}) \\
    &= \mathbb{E}_{(z_h,z_l),t,\epsilon} \left[ -(\log p_{\eta} (\hat{z}_{t_0}) - \log p_{\theta}(\hat{z}_{t_0})) \right]
\end{split}
\end{equation}
Instead of computing the intractable probability densities, the gradient of the DMD loss with respect to $\theta$ is estimated.
\begin{equation}
    \nabla_{\theta}\mathcal{L}_{DMD} = \mathbb{E}_{(z_h,z_l),t,\epsilon} \left[ -(s_{\eta} (\hat{z}_{t_0}) - s_{\theta}(\hat{z}_{t_0})) \nabla_{\theta} \hat{z}_{t_0} \right]
\end{equation}
where $s_{\eta}(\mathbf{x}) = \nabla_{\mathbf{x}} \log p_{\eta}(\mathbf{x})$ and $s_{\theta}(\mathbf{x}) = \nabla_{\mathbf{x}} \log p_{\theta}(\mathbf{x})$ represent the scores of each distribution. To leverage the property of the diffusion model, which provides scores of the diffused distribution, $\hat{z}_{t_0}$ is perturbed according to the timestep $t' \sim \mathcal{U}(0,1)$ and noise schedule of the  Teacher LDM. From this re-noised sample, each score can be obtained by utilizing the Teacher and Student LDMs, respectively \cite{scorebaseddiffusion}.

\noindent \textbf{Adversarial loss:} Flash Diffusion employs a diffusion-GAN hybrid methodology, which has demonstrated promising results in few-step data generation\cite{diffgan, ufogen}. After being perturbed by the teacher noise schedule and timestep $t'' \sim \mathcal{U}\{ 0.01, 0.25, 0.5, 0.75 \}$, $z_h$ and $\hat{z}_{t_0}$ are fed into the frozen  Teacher LDM. The discriminator $d_{\nu}$ is trained to distinguish the outputs of the  Teacher LDM. By leveraging the rich information from the  Teacher LDM, $d_{\nu}$ can employ a relatively simple architecture consisting of three convolutional layers with SiLU activation \cite{silu} and group normalization \cite{groupnorm}. Unlike the original Flash Diffusion, which uses only the encoder portion of the  Teacher LDM, we utilize the entire model for the discrimination process. The adversarial loss and discriminator loss are expressed as follows:

\begin{equation}
\begin{aligned}
    \mathcal{L}_{adv} &= \mathbb{E}_{(z_h,z_l),t_{i},t'',\epsilon} \left[ \lVert d_{\nu}(f_{\eta}(\hat{z}_{t''}, t'', z_l)) - 1 \rVert^{2} \right] \\
    \mathcal{L}_{disc} &= \frac{1}{2} \mathbb{E}_{(z_h,z_l),t_{i},t'',\epsilon} \left[ \lVert d_{\nu}(f_{\eta}(\hat{z}_{t''}, t'', z_l))\rVert^{2} \right.\\
    &\left. + \lVert d_{\nu}(f_{\eta}(z_{t''}, t'', z_l)) - 1  \rVert^{2} \right]
\end{aligned}
\end{equation}
where $z_{t''}$ and $\hat{z}_{t''}$ represent perturbed samples of $z_h$ and $\hat{z}_{t_0}$, respectively.

\subsection{SR Vocoder}
\label{subsec:srvocoder}

SR Vocoder generates a high-resolution waveform by conditioning on both the mel-spectrogram and the low-resolution input waveform. By integrating waveform-level information, it bypasses the need for LFR post-processing. As shown in Figure \ref{fig:overallarchitecture} (b), SR Vocoder consists of the LR Encoder and the Generator. For the Generator, we choose BigVGAN \cite{bigvgan}, which effectively produces waveforms without high-frequency artifacts by utilizing the anti-aliased multi-periodicity composition (AMP) module. We construct the LR Encoder using 1D convolutional layers and Leaky ReLU activation \cite{leakyrelu}. Each layer is implemented to match the output feature size of the corresponding layer in the Generator. The features from the LR Encoder are integrated with the Generator's features before the AMP module in each layer.

The training process of the SR Vocoder follows the BigVGAN-v2 configuration. SR Vocoder is trained using multi-scale mel spectrogram loss \cite{multimelloss}, adversarial loss, and feature matching loss \cite{featurematchingloss}. The Multi-Period Discriminator (MPD) \cite{hifigan} and Multi-Scale Sub-Band CQT (MS-SB-CQT) Discriminator \cite{cqtloss} are used for discrimination. We found that the SR Vocoder trained with MS-SB-CQT outperforms the SR Vocoder trained with the multi-resolution discriminator (MRD) \cite{mrd}, which is used in the original BigVGAN.

\section{Experiments}
\label{sec:experiments}

\begin{table*}[t]
\caption{Objective evaluation results, LSD/STFT-D}
\label{table:objectice}
\centering
\resizebox{\textwidth}{!}{
\begin{tabular}{llllllllll}
 & \multicolumn{3}{l}{\textbf{Speech}}&\multicolumn{3}{l}{\textbf{Music}}&\multicolumn{3}{l}{\textbf{Sound Effect}}\\ \Xhline{3\arrayrulewidth}
     
\multicolumn{1}{c|}{Method} &\multicolumn{1}{c}{4kHz}   &\multicolumn{1}{c}{8kHz}   &\multicolumn{1}{c|}{12kHz} &\multicolumn{1}{c}{4kHz}   &\multicolumn{1}{c}{8kHz}   &\multicolumn{1}{c|}{12kHz} &\multicolumn{1}{c}{4kHz}   &\multicolumn{1}{c}{8kHz}   &\multicolumn{1}{c}{12kHz} \\ \Xhline{3\arrayrulewidth}


\multicolumn{1}{c|}{Unprocessed}   &\multicolumn{1}{c}{3.05 / 1.33} &\multicolumn{1}{c}{2.68 / 1.05}&\multicolumn{1}{c|}{2.30 / 0.79} &\multicolumn{1}{c}{3.79 / 1.51} &\multicolumn{1}{c}{2.79 / 0.97}&\multicolumn{1}{c|}{1.95 / 0.58} &\multicolumn{1}{c}{3.72 / 1.47} &\multicolumn{1}{c}{2.77 / 0.95}&\multicolumn{1}{c}{1.96 / 0.57}\\

\multicolumn{1}{c|}{HiFi-GAN \cite{hifigan} (GT-Mel)}   &\multicolumn{1}{c}{0.72 / 0.27} &\multicolumn{1}{c}{0.72 / 0.27}&\multicolumn{1}{c|}{0.72 / 0.27} &\multicolumn{1}{c}{0.81 / 0.30} &\multicolumn{1}{c}{0.81 / 0.30}&\multicolumn{1}{c|}{0.81 / 0.30} &\multicolumn{1}{c}{0.79 / 0.29} &\multicolumn{1}{c}{0.79 / 0.29}&\multicolumn{1}{c}{0.79 / 0.29}\\

\multicolumn{1}{c|}{HiFi-GAN, LFR (GT-Mel)}   &\multicolumn{1}{c}{0.70 / 0.25} &\multicolumn{1}{c}{0.66 / 0.23}&\multicolumn{1}{c|}{0.61 / 0.20} &\multicolumn{1}{c}{0.84 / 0.30} &\multicolumn{1}{c}{0.79 / 0.27}&\multicolumn{1}{c|}{0.73 / 0.24} &\multicolumn{1}{c}{0.80 / 0.29} &\multicolumn{1}{c}{0.76 / 0.26}&\multicolumn{1}{c}{0.72 / 0.23}\\

\multicolumn{1}{c|}{SR Vocoder (GT-Mel)}   &\multicolumn{1}{c}{0.74 / 0.25} &\multicolumn{1}{c}{0.67 / 0.21}&\multicolumn{1}{c|}{0.59 / 0.16} &\multicolumn{1}{c}{0.88 / 0.31} &\multicolumn{1}{c}{0.80 / 0.25}&\multicolumn{1}{c|}{0.72 / 0.20} &\multicolumn{1}{c}{0.83 / 0.29} &\multicolumn{1}{c}{0.76 / 0.24}&\multicolumn{1}{c}{0.70 / 0.19}\\

\hline

\multicolumn{1}{c|}{NVSR-ResUNet \cite{nvsr}}   &\multicolumn{1}{c}{\textbf{1.53} / \textbf{0.57}} &\multicolumn{1}{c}{1.36 / 0.49}&\multicolumn{1}{c|}{1.18 / 0.42} &\multicolumn{1}{c}{2.40 / 1.00} &\multicolumn{1}{c}{2.38 / 0.96} &\multicolumn{1}{c|}{2.33 / 0.92} &\multicolumn{1}{c}{2.09 / 0.86}&\multicolumn{1}{c}{\textbf{1.95} / 0.76} &\multicolumn{1}{c}{2.01 / 0.76}\\

\multicolumn{1}{c|}{AudioSR \cite{audiosr} (100 NFEs)}   &\multicolumn{1}{c}{1.67 / 0.63} &\multicolumn{1}{c}{1.34 / 0.47}&\multicolumn{1}{c|}{0.95 / \textbf{0.31}} &\multicolumn{1}{c}{2.01 / 0.75} &\multicolumn{1}{c}{1.85 / 0.63}&\multicolumn{1}{c|}{1.86 / 0.59} &\multicolumn{1}{c}{2.02 / 0.74} &\multicolumn{1}{c}{2.00 / \textbf{0.67}}&\multicolumn{1}{c}{1.87 / \textbf{0.58}}\\

\multicolumn{1}{c|}{AudioSR (8 NFEs)}   &\multicolumn{1}{c}{1.81 / 0.74} &\multicolumn{1}{c}{1.79 / 0.69}&\multicolumn{1}{c|}{2.04 / 0.75} &\multicolumn{1}{c}{3.79 / 1.60} &\multicolumn{1}{c}{3.66 / 1.48}&\multicolumn{1}{c|}{3.61 / 1.39} &\multicolumn{1}{c}{3.27 / 1.32} &\multicolumn{1}{c}{3.81 / 1.50}&\multicolumn{1}{c}{3.89 / 1.45}\\

\hline

\multicolumn{1}{c|}{Student LDM (Proposed)}   &\multicolumn{1}{c}{1.65 / 0.63} &\multicolumn{1}{c}{1.34 / 0.49}&\multicolumn{1}{c|}{0.97 / 0.33} &\multicolumn{1}{c}{1.84 / 0.70} &\multicolumn{1}{c}{1.77 / 0.63}&\multicolumn{1}{c|}{1.51 / 0.51} &\multicolumn{1}{c}{1.92 / 0.71} &\multicolumn{1}{c}{2.09 / 0.73}&\multicolumn{1}{c}{1.83 / 0.60}\\

\multicolumn{1}{c|}{FlashSR (Proposed)}   &\multicolumn{1}{c}{1.62 / 0.62} &\multicolumn{1}{c}{\textbf{1.30} / \textbf{0.46}}&\multicolumn{1}{c|}{\textbf{0.94} / 0.32} &\multicolumn{1}{c}{\textbf{1.78} / \textbf{0.67}} &\multicolumn{1}{c}{\textbf{1.70} / \textbf{0.58}}&\multicolumn{1}{c|}{\textbf{1.45} / \textbf{0.47}} &\multicolumn{1}{c}{\textbf{1.88} / \textbf{0.69}} &\multicolumn{1}{c}{2.06 / 0.70}&\multicolumn{1}{c}{\textbf{1.80} / \textbf{0.58}}\\

\Xhline{3\arrayrulewidth}

\end{tabular}
}
\end{table*}

\begin{figure*}[!t]
\centering
\begin{minipage}{0.8\textwidth}
  \centering
  \centerline{\includegraphics[width=\textwidth]{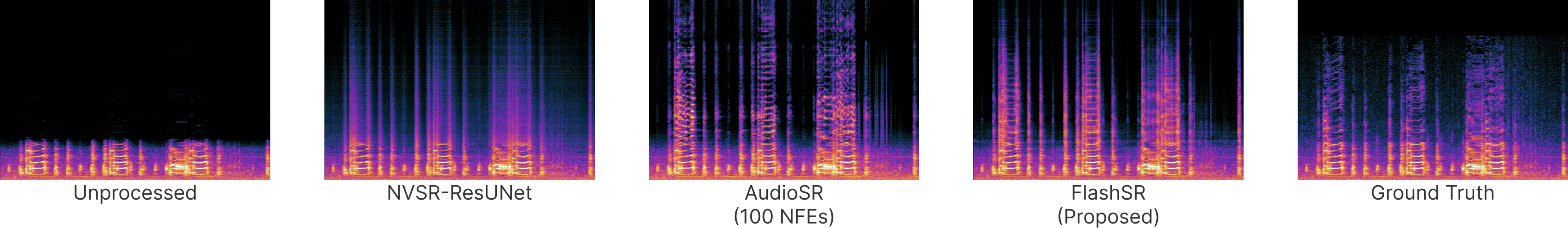}}
\end{minipage}
\caption{Spectrogram of compared models.}
\label{fig:spec}
\end{figure*}

\subsection{Training Dataset and Preprocessing}
We follow the training dataset configuration and data simulation method of AudioSR \cite{audiosr}.  Student LDM and SR Vocoder are optimized using the  OpenSLR\footnote{https://openslr.org/} speech dataset \cite{voicefixer}, MedleyDB \cite{medleydb}, MUSDB18-HQ \cite{musdb18}, MoisesDB \cite{moisedb}, and FreeSound \cite{wavcaps}\footnote{https://labs.freesound.org/}. The total duration of the audio is approximately 3,800 hours. All audio sample rates were resampled to 48kHz and randomly segmented to match the length of 5.12 seconds. The mel-spectrogram is extracted from the audio with the Hann window size of 2048, the hop size of 480, and the 256-channel mel filterbank. The channel size and compression level of VAE latent are set to 16 and 8, respectively. To simulate low-resolution audio, various cutoff frequencies, lowpass filter types, and filter orders are dynamically selected during the training process. Cutoff frequencies are sampled from $\mathcal{U}\{2000, 16000\}$, filter types from $\mathcal{U}\{Chebyshev, Butterworth, Bessel, Elliptic\}$, and filter orders from $\mathcal{U}\{2, 10\}$. 
\subsection{Implementation Details}
Of the 3,800 hours of training data, around 3,000 hours are from the Freesound dataset. This imbalance in sound categories may lead to overfitting. To prevent potential issues, we sampled the audio to ensure an even distribution of sound categories in each mini-batch. 

The training procedure for Student LDM follows \cite{flashdiffusion} unless specified. Student LDM and the discriminator were trained using the AdamW optimizer with a batch size of 16 and a learning rate of $10^{-5}$ for 30K steps. 
The loss weights $\lambda_{adv}$ and $\lambda_{dmd}$ start at 0, increasing uniformly every 5K steps until 20K steps, reaching final values of 0.3 and 0.7, respectively. We use the official checkpoint of LDM and the VAE from AudioSR\footnote{https://github.com/haoheliu/versatile\_audio\_super\_resolution}, which is used as Teacher LDM. To generate the target for the distillation loss, we utilize DPM Solver++ \cite{dpmsolverplusplus} 
and set the guidance scale to 4. Empirically, we found that the optimal guidance scale is slightly higher than the one used for sampling with the Teacher LDM. The Student LDM has approximately 258M parameters in total. During training, only about 45M parameters were updated by applying LoRA \cite{lora} to the attention modules. 

SR Vocoder, MPD, and MS-SB-CQT were trained using the AdamW optimizer with a batch size of 4 for 2.5M steps. The learning rate was set to $5 \times 10^{-5}$ for the SR Vocoder and $10^{-4}$ for MPD and MS-SB-CQT. The learning rates are decreased at each training step by multiplying the previous value by 0.9999996.
\subsection{Evaluation}
\noindent \textbf{Comparison:} We compare FlashSR (1 NFE) with AudioSR using 50 steps (100 NFEs) and 4 steps (8 NFEs), both with CFG. We use the official checkpoint and implementation of AudioSR. The guidance scale for AudioSR was set to 3.5, and DPM Solver++ \cite{dpmsolverplusplus} was used for sampling. We also compare our model with NVSR \cite{nvsr}, originally designed for speech super-resolution. We trained the ResUNet of NVSR using the same dataset as FlashSR. Since NVSR performs upsampling at the mel-spectrogram level, it also requires a neural vocoder to generate the final waveform output.  HiFi-GAN and LFR postprocessing were used for all comparison models. To assess the effectiveness of SR Vocoder, we also evaluated the Student LDM using HiFi-GAN and LFR.

\noindent \textbf{Dataset:} We use the VCTK test set (speech) \cite{nvsr}, FMA-small (music) \cite{fmasmall}, and ESC50 fold-5 (sound effects) \cite{esc50} for both objective and subjective evaluations. We selected 100 music samples from various genres in the FMA-small and 200 samples from ESC50 fold-5.

\noindent \textbf{Evaluation:} For the objective evaluation, we evaluate each sound category using three different cutoff frequencies: 4kHz, 8kHz, and 12kHz. We used two evaluation metrics: Log-Spectral Distance (LSD) with the base-10 logarithm, as used in previous studies \cite{audiosr, li2015dnn, nvsr}, and Spectrogram L1 Distance (STFT-D). For diffusion-based methods, we additionally evaluate the real-time factor (RTF) for generating 5.12 seconds of audio on a single A6000 GPU. For the subjective evaluation, we conducted a listening test with 21 participants for each sound category. The cutoff frequency was set to 4kHz for all cases. Participants were given the low-resolution input audio as a low anchor and rated the overall quality on a scale from 1 to 5. 
We evaluated NVSR-ResUNet, AudioSR (100 NFEs), and the proposed FlashSR.

\section{Results}

\begin{figure}[!t]
\centering
\begin{minipage}{0.4\textwidth}
  \centering
  \centerline{\includegraphics[width=\textwidth]{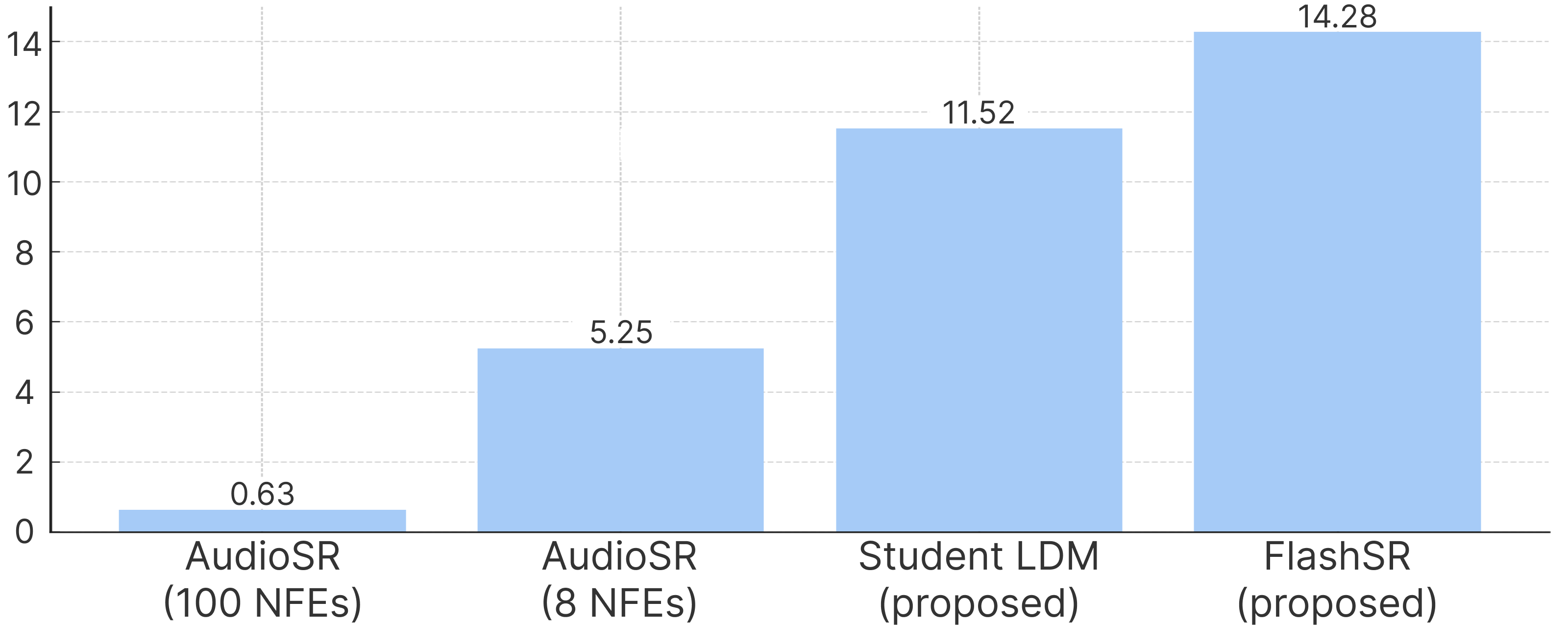}}
\end{minipage}
\caption{Inverse real time factor (RTF)}
\label{fig:rtf}
\end{figure}

\subsection{Objective Evaluation}
Table \ref{table:objectice} shows the objective evaluation results. While FlashSR demonstrates significantly faster inference speed than AudioSR (100 NFEs), as shown in Figure \ref{fig:rtf}, it also achieves comparable or superior results across all test cases. In the speech SR task, FlashSR outperforms AudioSR (100 NFEs) in all metrics and cases, except for STFT-D when the cutoff frequency is set to 12kHz. In the music SR task, FlashSR achieves state-of-the-art performance across every cutoff frequency setting. In the sound effect SR task, FlashSR surpasses AudioSR (100 NFEs) at the 4kHz and 12kHz cutoff frequency settings. In all test cases, FlashSR shows superior performance compared to Student LDM. This indicates that the SR Vocoder can more effectively produce high-resolution audio from generated mel-spectrograms than HiFi-GAN with LFR. In all test cases, AudioSR (8 NFEs) performs the worst among the comparisons, indicating that reducing sampling steps without distillation leads to poor generation quality.

\subsection{Subjective Evaluation}
Table \ref{table:subjective} shows the results of the subjective evaluation. FlashSR attains the best scores across all test cases. Although NVSR-ResUNet performs best at the 4kHz cutoff frequency setting in the speech SR task, we observed that LSD does not always align with perceptual quality, as noted in \cite{audiosr}. Contrary to the objective evaluation results, NVSR-ResUNet achieves lower scores than unprocessed low-resolution audio in both the speech SR and sound effect SR tasks. We observed that the audio samples generated by NVSR-ResUNet were noisy and showed a blurred harmonic structure, as shown in Figure \ref{fig:spec}. We also found that while AudioSR performs upsampling robustly, it tends to produce excessive sibilance.

\begin{table}[t]
\caption{Subjective evaluation results (4kHz cutoff frequency).}
\label{table:subjective}
\centering
\footnotesize{
\begin{tabular}{llll}

\Xhline{3\arrayrulewidth}

\multicolumn{1}{c|}{Method} &\multicolumn{1}{c}{Speech}   &\multicolumn{1}{c}{Music}   &\multicolumn{1}{c}{Sound Effect} \\ \Xhline{3\arrayrulewidth}

\multicolumn{1}{c|}{Unprocessed} & \multicolumn{1}{c}{2.22} & \multicolumn{1}{c}{2.11} & \multicolumn{1}{c}{2.43}\\

\multicolumn{1}{c|}{Ground Truth} & \multicolumn{1}{c}{4.51} & \multicolumn{1}{c}{3.80} & \multicolumn{1}{c}{4.10}\\

\hline

\multicolumn{1}{c|}{NVSR-ResUNet \cite{nvsr}} & \multicolumn{1}{c}{2.04} & \multicolumn{1}{c}{2.21} & \multicolumn{1}{c}{2.41}\\

\multicolumn{1}{c|}{AudioSR \cite{audiosr} (100 NFEs)} & \multicolumn{1}{c}{3.40} & \multicolumn{1}{c}{3.28} & \multicolumn{1}{c}{3.12}\\

\multicolumn{1}{c|}{FlashSR} & \multicolumn{1}{c}{\textbf{3.67}} & \multicolumn{1}{c}{\textbf{3.56}} & \multicolumn{1}{c}{\textbf{3.77}}\\

\Xhline{3\arrayrulewidth}

\end{tabular}
}

\end{table}

\section{Conclusions}
We propose FlashSR, a fast and effective diffusion model for versatile audio super-resolution. To accelerate inference speed without compromising audio quality, we explore diffusion distillation, previously unexplored in audio super-resolution tasks. We also propose the SR Vocoder that generates the high-resolution waveform from the mel-spectrogram and low-resolution waveform with the arbitrary sample rate. FlashSR is about 22 times faster than AudioSR, while it achieves comparable generation quality in both objective and subjective evaluations. Future work involves enhancing audio quality by improving the Teacher LDM and fine-tuning the Vocoder with generated mel-spectrograms.

\vfill\pagebreak

\bibliographystyle{ieeetr.bst}
\bibliography{IEEEabrv.bib}

\begin{thebibliography}{10}

\bibitem{audiosr}
H.~Liu, K.~Chen, Q.~Tian, W.~Wang, and M.~D. Plumbley, ``Audiosr: Versatile audio super-resolution at scale,'' in {\em ICASSP}, pp.~1076--1080, IEEE, 2024.

\bibitem{li2015dnn}
K.~Li, Z.~Huang, Y.~Xu, and C.-H. Lee, ``Dnn-based speech bandwidth expansion and its application to adding high-frequency missing features for automatic speech recognition of narrowband speech.,'' in {\em INTERSPEECH}, pp.~2578--2582, 2015.

\bibitem{hou2020}
N.~Hou, C.~Xu, J.~T.~Z. Van Tung~Pham, E.~S. Chng, and H.~Li, ``Speaker and phoneme-aware speech bandwidth extension with residual dual-path network,'' in {\em INTERSPEECH}, pp.~4064--4068, 2020.

\bibitem{nuwave}
J.~Lee and S.~Han, ``Nu-wave: A diffusion probabilistic model for neural audio upsampling,'' in {\em INTERSPEECH}, pp.~1634--1638, 2021.

\bibitem{bwallyou}
J.~Su, Y.~Wang, A.~Finkelstein, and Z.~Jin, ``Bandwidth extension is all you need,'' in {\em ICASSP}, pp.~696--700, IEEE, 2021.

\bibitem{musicsr}
S.~Hu, B.~Zhang, B.~Liang, E.~Zhao, and S.~Lui, ``Phase-aware music super-resolution using generative adversarial networks.,'' in {\em INTERSPEECH}, pp.~4074--4078, 2020.

\bibitem{stablediffusion}
R.~Rombach, A.~Blattmann, D.~Lorenz, P.~Esser, and B.~Ommer, ``High-resolution image synthesis with latent diffusion models,'' in {\em CVPR}, pp.~10684--10695, 2022.

\bibitem{vae}
D.~P. Kingma and M.~Welling, ``Auto-encoding variational bayes,'' in {\em ICLR}, 2014.

\bibitem{hifigan}
J.~Kong, J.~Kim, and J.~Bae, ``Hifi-gan: Generative adversarial networks for efficient and high fidelity speech synthesis,'' {\em NeurIPS}, pp.~17022--17033, 2020.

\bibitem{ddim}
J.~Song, C.~Meng, and S.~Ermon, ``Denoising diffusion implicit models,'' in {\em ICLR}, 2021.

\bibitem{cfg}
J.~Ho and T.~Salimans, ``Classifier-free diffusion guidance,'' in {\em NeurIPS 2021 Workshop on Deep Generative Models and Downstream Applications}, 2021.

\bibitem{dpmsolverplusplus}
C.~Lu, Y.~Zhou, F.~Bao, J.~Chen, C.~Li, and J.~Zhu, ``Dpm-solver++: Fast solver for guided sampling of diffusion probabilistic models,'' {\em arXiv preprint arXiv:2211.01095}, 2022.

\bibitem{progressivedistill}
T.~Salimans and J.~Ho, ``Progressive distillation for fast sampling of diffusion models,'' in {\em ICLR}, 2022.

\bibitem{flashdiffusion}
C.~Chadebec, O.~Tasar, E.~Benaroche, and B.~Aubin, ``Flash diffusion: Accelerating any conditional diffusion model for few steps image generation,'' {\em arXiv preprint arXiv:2406.02347}, 2024.

\bibitem{consistencyTTA}
Y.~Bai, T.~Dang, D.~Tran, K.~Koishida, and S.~Sojoudi, ``Accelerating diffusion-based text-to-audio generation with consistency distillation,'' {\em arXiv preprint arXiv:2309.10740}, 2023.

\bibitem{musicconsistency}
Z.~Fei, M.~Fan, and J.~Huang, ``Music consistency models,'' {\em arXiv preprint arXiv:2404.13358}, 2024.

\bibitem{liu2024audiolcm}
H.~Liu, R.~Huang, Y.~Liu, H.~Cao, J.~Wang, X.~Cheng, S.~Zheng, and Z.~Zhao, ``Audiolcm: Text-to-audio generation with latent consistency models,'' {\em arXiv preprint arXiv:2406.00356}, 2024.

\bibitem{soundctm}
K.~Saito, D.~Kim, T.~Shibuya, C.-H. Lai, Z.~Zhong, Y.~Takida, and Y.~Mitsufuji, ``Soundctm: Uniting score-based and consistency models for text-to-sound generation,'' {\em arXiv preprint arXiv:2405.18503}, 2024.

\bibitem{prodiff}
R.~Huang, Z.~Zhao, H.~Liu, J.~Liu, C.~Cui, and Y.~Ren, ``Prodiff: Progressive fast diffusion model for high-quality text-to-speech,'' in {\em ACM-MM}, 2022.

\bibitem{comospeech}
Z.~Ye, W.~Xue, X.~Tan, J.~Chen, Q.~Liu, and Y.~Guo, ``Comospeech: One-step speech and singing voice synthesis via consistency model,'' in {\em ACM-MM}, pp.~1831--1839, 2023.

\bibitem{lora}
E.~J. Hu, Y.~Shen, P.~Wallis, Z.~Allen-Zhu, Y.~Li, S.~Wang, L.~Wang, and W.~Chen, ``Lora: Low-rank adaptation of large language models.,'' in {\em ICLR}, 2022.

\bibitem{ddpm}
J.~Ho, A.~Jain, and P.~Abbeel, ``Denoising diffusion probabilistic models,'' in {\em Advances in Neural Information Processing Systems}, vol.~33, pp.~6840--6851, 2020.

\bibitem{dmdloss}
T.~Yin, M.~Gharbi, R.~Zhang, E.~Shechtman, F.~Durand, W.~T. Freeman, and T.~Park, ``One-step diffusion with distribution matching distillation,'' in {\em CVPR}, pp.~6613--6623, 2024.

\bibitem{scorebaseddiffusion}
Y.~Song, J.~Sohl{-}Dickstein, D.~P. Kingma, A.~Kumar, S.~Ermon, and B.~Poole, ``Score-based generative modeling through stochastic differential equations,'' in {\em ICLR}, 2021.

\bibitem{diffgan}
Z.~Xiao, K.~Kreis, and A.~Vahdat, ``Tackling the generative learning trilemma with denoising diffusion {GAN}s,'' in {\em ICLR}, 2022.

\bibitem{ufogen}
Y.~Xu, Y.~Zhao, Z.~Xiao, and T.~Hou, ``Ufogen: You forward once large scale text-to-image generation via diffusion gans,'' in {\em CVPR}, pp.~8196--8206, 2024.

\bibitem{silu}
D.~Hendrycks and K.~Gimpel, ``Gaussian error linear units (gelus),'' {\em arXiv preprint arXiv:1606.08415}, 2016.

\bibitem{groupnorm}
Y.~Wu and K.~He, ``Group normalization,'' in {\em ECCV}, pp.~3--19, 2018.

\bibitem{bigvgan}
S.~gil Lee, W.~Ping, B.~Ginsburg, B.~Catanzaro, and S.~Yoon, ``Big{VGAN}: A universal neural vocoder with large-scale training,'' in {\em ICLR}, 2023.

\bibitem{leakyrelu}
A.~Maas, A.~Hannun, and A.~Ng, ``Rectifier nonlinearities improve neural network acoustic models,'' in {\em ICML}, 2013.

\bibitem{multimelloss}
R.~Kumar, P.~Seetharaman, A.~Luebs, I.~Kumar, and K.~Kumar, ``High-fidelity audio compression with improved rvqgan,'' {\em NeurIPS}, vol.~36, 2024.

\bibitem{featurematchingloss}
A.~B.~L. Larsen, S.~K. S{\o}nderby, H.~Larochelle, and O.~Winther, ``Autoencoding beyond pixels using a learned similarity metric,'' in {\em ICML}, pp.~1558--1566, 2016.

\bibitem{cqtloss}
Y.~Gu, X.~Zhang, L.~Xue, and Z.~Wu, ``Multi-scale sub-band constant-q transform discriminator for high-fidelity vocoder,'' in {\em ICASSP}, pp.~10616--10620, 2024.

\bibitem{mrd}
W.~Jang, D.~Lim, J.~Yoon, B.~Kim, and J.~Kim, ``{UnivNet: A Neural Vocoder with Multi-Resolution Spectrogram Discriminators for High-Fidelity Waveform Generation},'' in {\em INTERSPEECH}, pp.~2207--2211, 2021.

\bibitem{nvsr}
H.~Liu, W.~Choi, X.~Liu, Q.~Kong, Q.~Tian, and D.~Wang, ``Neural vocoder is all you need for speech superresolution,'' in {\em INTERSPEECH}, p.~4227–4231, 2022.

\bibitem{voicefixer}
H.~Liu, Q.~Kong, Q.~Tian, Y.~Zhao, D.~Wang, C.~Huang, and Y.~Wang, ``Voicefixer: Toward general speech restoration with neural vocoder,'' {\em arXiv preprint arXiv:2109.13731}, 2021.

\bibitem{medleydb}
R.~M. Bittner, J.~Salamon, M.~Tierney, M.~Mauch, C.~Cannam, and J.~P. Bello, ``Medleydb: {A} multitrack dataset for annotation-intensive {MIR} research,'' in {\em ISMIR}, pp.~155--160, 2014.

\bibitem{musdb18}
Z.~Rafii, A.~Liutkus, F.-R. St{\"o}ter, S.~I. Mimilakis, and R.~Bittner, ``The musdb18 corpus for music separation,'' 2017.

\bibitem{moisedb}
I.~Pereira, F.~Ara{\'{u}}jo, F.~Korzeniowski, and R.~Vogl, ``Moisesdb: {A} dataset for source separation beyond 4-stems,'' in {\em ISMIR}, pp.~619--626, 2023.

\bibitem{wavcaps}
X.~Mei, C.~Meng, H.~Liu, Q.~Kong, T.~Ko, C.~Zhao, M.~D. Plumbley, Y.~Zou, and W.~Wang, ``Wavcaps: A chatgpt-assisted weakly-labelled audio captioning dataset for audio-language multimodal research,'' {\em TASLP}, 2024.

\bibitem{fmasmall}
M.~Defferrard, K.~Benzi, P.~Vandergheynst, and X.~Bresson, ``{FMA:} {A} dataset for music analysis,'' in {\em ISMIR}, pp.~316--323, 2017.

\bibitem{esc50}
K.~J. Piczak, ``{ESC:} dataset for environmental sound classification,'' in {\em ACM}, pp.~1015--1018, 2015.

\end{thebibliography}

\end{document}